\documentclass[aps,prb,floatfix,showpacs,twocolumn]{revtex4}
\usepackage{amsmath,amssymb}
\usepackage{epsfig}

\newcommand{\bm}{\boldsymbol}

\begin{document}

\hsize\textwidth\columnwidth\hsize\csname@twocolumnfalse\endcsname

\title{Coulomb Drag in Graphene}

\author{Wang-Kong Tse$^1$}
\author{Ben Yu-Kuang Hu$^{1,2}$}
\author{S. Das Sarma$^1$}
\affiliation{$^1$Condensed Matter Theory Center, Department of
Physics, University of Maryland, College Park, Maryland 20742}
\affiliation{$^2$Department of Physics, University of Akron, Akron, Ohio 44325-4001}

\begin{abstract}
We study the Coulomb drag between two single graphene sheets in intrinsic and extrinsic 
graphene systems with no interlayer tunneling. 
The general expression for the nonlinear susceptibility appropriate for single-layer graphene systems 
is derived using the diagrammatic perturbation theory, and the corresponding exact zero-temperature expression is 
obtained analytically. We find that, despite the existence of a non-zero  
conductivity in an intrinsic graphene layer, the Coulomb drag between intrinsic graphene 
layers vanishes at all temperatures. In extrinsic systems, we obtain  
numerical results and an approximate analytical result 
for the drag resistivity $\rho_{\textrm{D}}$, and find that $\rho_{\textrm{D}}$ 
goes as $T^2$ at low temperature $T$, as $1/d^4$ for large bilayer 
separation $d$ and $1/n^3$ for high carrier density $n$. 
We also discuss qualitatively the effect of plasmon-induced 
enhancement on the Coulomb drag, which should occur 
at a temperature of the order of or higher than the Fermi temperature. 
\end{abstract}

\maketitle
\newpage

\section{INTRODUCTION}
With the recent advent of the experimental fabrication of a single
layer of graphene, the electronic and transport properties of this
newly discovered material have been intensively studied both 
experimentally \cite{exp1,spectral} and theoretically
\cite{Polar,gFL}. Whereas electronic structure experiments have revealed
detailed subtle many-body effects on the graphene energy spectrum, 
transport experiments have also revealed some apparently unusual
features of graphene transport properties, most noticably, that the
conductivity has a non-zero minimum value around zero bias gate voltage. 
Up to now, the transport experiments performed have been focused only 
on the longitudinal and Hall transport properties 
(in both weak and strong magnetic fields, including the quantum Hall
regime) and weak localization, where all of these phenomena depend 
essentially only on the physics of scattering of individual single 
quasiparticle from impurities with electron-electron many-body 
interaction effect playing the role of a small quantitative correction.  
In two-dimensional electron gas (2DEG) semiconductor bilayer
structures (e.g. modulation-doped GaAs/Al$_x$Ga$_{1-x}$As double
quantum wells), electron-electron scattering between the 2DEG layers
give rise to the effect of Coulomb drag where a `drag' 
current is induced purely from the momentum exchanges through interlayer
electron-electron scattering events. One measures the effect of 
Coulomb drag by the drag resistivity $\rho_{\textrm{D}}$, which is defined by the induced
drag electric field in the open-circuited passive layer per unit applied
current density in the active layer. In high-mobility samples where the
disorder is weak, $\rho_{\textrm{D}}$ goes as $T^2$ at low temperatures $T$, and
as $1/d^4$ for large bilayer separation $d$ (Refs.~\onlinecite{Gramila,Smith}). 

In this paper, we investigate the Coulomb drag in graphene ``bilayer''
systems with no interlayer tunneling, considering both the intrinsic (chemical potential $\mu = 0$) and extrinsic ($\mu \neq 0$) cases. We emphasize right in the beginning so that there is no semantic confusion what we mean by the terminology `bilayer' graphene. Our `bilayer' graphene is two isolated parallel 2D graphene monolayers separated by a distance $d$, with \textit{no} interlayer tunneling. The electronic structure of each graphene monolayer is thus unaffected by having the other layer. Each graphene layer is assumed to have its own variable carrier density in the extrinsic case. Our system is thus different from the ordinary bilayer graphene where $d \sim 1-5\AA$ with strong interlayer tunneling. 
Throughout this paper, we shall also use the terms 
``undoped'' and ``doped'' interchangeably with ``intrinsic'' and ``extrinsic'' respectively; keeping in mind that 
in experiments the chemical potential can be changed by both chemical doping and gating with an applied voltage. 
The Coulomb drag in graphene is interesting not only because it is a novel material with a 
linear energy spectrum, but also because it only spans a thickness of 
a single carbon atom, the electrons are much more confined along the
perpendicular direction comparing with 2DEG in a quantum well, where
the finite width thickness has to be taken into account in any quantitative
comparison with experiments. Thus, the Coulomb drag phenomenon in
graphene is expected to be theoretically very well accounted for with 
two zero-thickness graphene sheets. In addition, tunneling is only appreciable when 
the the out-of-plane $\pi$ orbitals from the two graphene sheets start to overlap with each other at an interlayer distance $d$ of a few angstroms ($d \simeq 3.5 \AA$ in naturally occuring graphite), making it possible to study the effect of Coulomb drag for an interlayer separation $d$ down to a few tens of angstroms, about an order of magnitude smaller than is possible in the usual double quantum well systems without appreciable tunneling. 


\section{FORMALISM}
Graphene has a real-space honeycomb lattice structure with two
interpenetrating sublattices, giving a honeycomb reciprocal space
structure with the Brillouin zone forming a hexagon. At each of
these corners where the high symmetry point is denoted as K or
K', the energy dispersion is linear and the effective low-energy
Hamiltonian at the vicinity of these K and K' points is given by
$H_0 = v\bm{\hat{\sigma}}\cdot\bm{k}$. The eigenenergy is
$\epsilon_{\bm{k}\lambda} = \lambda vk$, where the chirality label
$\lambda = \pm 1$ describes conduction band ($\lambda = 1$) and
valence band ($\lambda = -1$) spectra. Throughout this paper we choose $\hbar = 1$ 
unless it is written out explicitly. The transformation which
diagonalizes the Hamiltonian is a local transformation, dependent
on the momentum $\bm{k}$ through $\mathrm{tan}\phi_{\bm{k}} =
k_y/k_x$:
\begin{eqnarray}
U_{\bm{k}} = \frac{1}{\sqrt{2}} \left[
\begin{array}{cc}
1 & 1 \\
e^{i\phi_{\bm{k}}} & -e^{i\phi_{\bm{k}}}
\end{array}
\right]. \label{eq1}
\end{eqnarray}
%

The central quantity in the Coulomb drag problem is the nonlinear
susceptibility \cite{Oreg,Jauho,mydrag}, $\Gamma$:
%
\begin{figure}[t]
  \includegraphics[width=4.5cm,angle=0]{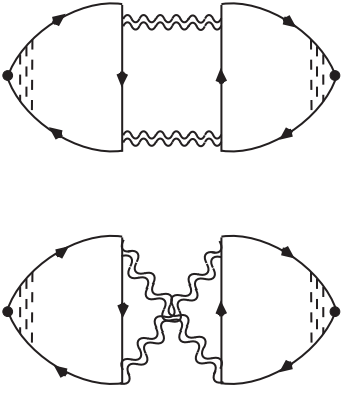}
\caption{Diagrams contributing to the drag resistivity
  Eq.~(\ref{eq3}). The double wavy lines represent the screened interlayer Coulomb potential Eq.~(\ref{eq62}) and
  the vertices on the left and on the right denote charge current in the two
layers. Dashed vertical lines next to the vertices denote impurity vertex correction to the charge current.} 
\label{fig1}
\end{figure}
\begin{eqnarray}
&&\bm{\Gamma}(\bm{q},\omega) = \int
\frac{\mathrm{d}\varepsilon}{2\pi i}
\left[n_F(\varepsilon+\omega)-n_F(\varepsilon)\right]\sum_k
\mathrm{tr} \nonumber \\
&&\;\left\{\left[G_{\bm{k}-\bm{q}}^{A}(\varepsilon)-G_{\bm{k}-\bm{q}}^{R}(\varepsilon)\right]
G_{\bm{k}}^{A}(\varepsilon+\omega)\bm{J}(\bm{k})G_{\bm{k}}^{R}(\varepsilon+\omega)\right\}
\nonumber \\
&&+\left\{\bm{q},\omega \to -\bm{q},-\omega\right\}, \label{eq2}
\end{eqnarray}
where $G_{\bm{k}}^{R,A}(\varepsilon) = (\varepsilon-H_0 \pm
i/2\tau)^{-1}$ denotes, within the Born approximation for the
self-energy, the impurity-averaged retarded/advanced Green function, $\tau$ the lifetime due to impurity scattering,
 $\bm{J}$ the charge current vertex, $n_F$ the Fermi function and `$\mathrm{tr}$' the trace. In the rest of this paper, we shall
simply denote the $x$-components of $\bm{\Gamma}$ and $\bm{J}$ as
$\Gamma$ and $J$.
The drag conductivity, diagrammatically shown in
Fig.~\ref{fig1}, is given by \cite{Oreg,Jauho,mydrag}
%
%
%
\begin{eqnarray}
\sigma_{\mathrm{D}} &=& \frac{1}{16\pi k_B
T}\sum_{q}\int_{0}^{\infty}
\frac{\mathrm{d}\omega}{\mathrm{sinh}^2\left(\hbar\omega/2k_B
    T\right)} \nonumber \\
&&\Gamma_{1}\left(q,\omega\right)\Gamma_{2}\left(q,\omega\right)\left\vert
U_{12}\left(q,\omega\right)\right\vert^2, \label{eq3}
\end{eqnarray}
here subscripts `1' and `2' are the labels for the two
single-layer graphene, $U_{12}$ is the screened interlayer
potential, which in the random phase approximation (RPA) is given by 
\begin{eqnarray}
&&U_{12}\left(q,\omega\right) = \label{eq62} \\
&&\frac{V(q)e^{-qd}}{[1+\Pi_1 V(q)][1+\Pi_2 V(q)]-\Pi_1\Pi_2
V^2(q)e^{-2qd}}, \nonumber
\end{eqnarray}
where $d$ is the interlayer spacing, $V(q) = 2\pi e^2/q$ is the 
bare Coulomb potential, $\Pi\left(q,\omega\right)$ is the graphene 
polarizability \cite{Polar}. Instead of the drag conductivity, in
experiments one measures the drag resistivity $\rho_{\textrm{D}}$ which is defined as (here
$W$ and $L$ are the width and length of the
sample, respectively) $\rho_{\textrm{D}} \equiv (W/L)(V_2/I_1)$, the ratio
of the induced voltage in the passive layer $V_2$ to the applied
current in the active layer $I_1$. The drag resistivity is then 
obtained from the drag conductivity as $\rho_{\textrm{D}} =
\sigma_{\textrm{D}}/(\sigma_{\textrm{L}1}\sigma_{\textrm{L}2}-\sigma_{\textrm{D}}^2) \simeq \sigma_{\textrm{D}}/\sigma_{\textrm{L}1}\sigma_{\textrm{L}2}$, where
$\sigma_{\textrm{L}1,2}$ is the longitudinal conductivity of the individual
layer $1$ or $2$.

In this paper, we shall restrict ourselves to the Boltzmann
regime ($\omega\tau
\gg 1$ or $ql \gg 1$, where $l = v\tau$ is the mean free path) corresponding
to weak impurity scattering, which is the case relevant to actual
experimental situations where high-mobility samples with dilute
impurities are used. The longitudinal current for the graphene Hamiltonian is $\bm{J} = e\partial H_0/\partial \bm{k} 
= ev\bm{\hat{\sigma}}$. In the presence of impurities, 
vertex correction to the current is taken into account within the impurity ladder 
approximation, which gives the impurity-dressed current vertex as $\bm{J} = (\tau_{\textrm{tr}}/\tau)ev\bm{\hat{\sigma}}$, 
where the transport time $\tau_{\textrm{tr}}$ for graphene is given by 
\begin{equation}
\tau_{\textrm{tr}}^{-1} = \pi\sum_{k'}n_i|u_i(\bm{k}-\bm{k}')|^2(1-\mathrm{cos}^2\theta_+)\delta(\epsilon_{\bm{k}+}-\epsilon_{\bm{k}'+})\bigg\vert_{k = k_F},
\label{eq12}
\end{equation}
with $n_i$ and $u_i$ being respectively the impurity density and impurity potential; $\theta_{+}
= \phi_{\bm{k+q}}-\phi_{\bm{k}}$ being the scattering angle from momentum $\bm{k}$ to $\bm{k+q}$. 
We have given Eq.~(\ref{eq12}) for the transport time here only for the purpose of completeness, 
as our final results for the drag resistivity do not depend on $\tau_{\textrm{tr}}$. 
%

Following Ref.~\onlinecite{mydrag}, we express the nonlinear
susceptibility Eq.~(\ref{eq2}) as
\begin{eqnarray}
&&{\Gamma}(\bm{q},\omega) = \tau
\sum_{\lambda,\lambda'=\pm}\sum_{k}
\left[\tilde{J}_{\lambda\lambda}(\bm{k}+\bm{q})-\tilde{J}_{\lambda'\lambda'}(\bm{k})\right] \label{eq4} \\
&&\mathrm{Im}\left\{\left(1+\lambda\lambda'\mathrm{cos}\theta_+\right)\frac{n_F(\epsilon_{\bm{k}\lambda'})-n_F(\epsilon_{\bm{k+q}\lambda})}{\omega+\epsilon_{\bm{k}\lambda'}-\epsilon_{\bm{k+q}\lambda}+i0^{+}}\right\}.
\nonumber
\end{eqnarray}
where $\tilde{{J}}({\bm{k}}) = U_{\bm{k}}^{\dag}{J}U_{\bm{k}}$ is
the impurity-dressed charge current vertex expressed in the chiral basis. 
Eq.~(\ref{eq4}) is different from the
nonlinear susceptibility for regular 2DEG with quadratic spectrum in
two ways: (1) there are contributions to the electron-hole excitations
coming from intraband transitions ($\lambda = \lambda'$) and interband
transitions ($\lambda \neq \lambda'$); (2) there is an 
additional factor $(1\pm\mathrm{cos}\theta_{+})/2$, which 
derives from the Berry phase structure of the graphene
Hamiltonian. Furthermore, the nonlinear susceptibility Eq.~(\ref{eq4}) is not directly 
proportional to the imaginary part of the polarizability as in regular 2DEG, 
because here the current $\tilde{\bm{J}}(\bm{k})$ is not directly proportional 
to the momentum $\bm{k}$. Eq.~(\ref{eq4}) has the same formal structure as in
the case of a regular 2DEG with Rashba/Dresselhaus spin-orbit
coupling \cite{mydrag}, where $\lambda = \pm 1$ describes the two
spin-split bands. The finite off-diagonal components of 
$\tilde{{J}}$ do not contribute to the nonlinear susceptibility,
and only the diagonal components $\tilde{{J}}_{\lambda\lambda} = 
\lambda (\tau_{\textrm{tr}}/\tau)e v\;\mathrm{cos}\phi_{\bm{k}}$ enter into the expression
Eq.~(\ref{eq4}), corresponding to electrons moving in the
conduction band ($\lambda = 1$) with a velocity of constant magnitude 
$(\tau_{\textrm{tr}}/\tau)\bm{v}$ and valence band ($\lambda = -1$) with $-(\tau_{\textrm{tr}}/\tau)\bm{v}$. In the 
following, we consider the Coulomb drag between intrinsic
graphene layers and extrinsic graphene layers separately.

\section{DRAG IN INTRINSIC GRAPHENE SYSTEMS}
For the case where the graphene layers are undoped 
$\mu = 0$, we first state the main result: the drag
conductivity between two intrinsic graphene layers, or between one extrinsic
and one intrinsic graphene layers, is identically zero. This is not at 
first sight a trivial consequence of zero doping if one recalls there 
is a finite conductivity (so-called the ``minimum conductivity'') at zero doping in graphene. 
A physical explanation and a general argument for the reason why 
this is so is in order. When the Fermi 
level is at the Dirac point, the only process for electron-hole 
pair creation will be interband electron excitation from the valence 
band to the conduction band by which equal numbers of electrons and holes 
are created. In the mechanism of Coulomb 
drag, the applied electric field drives the electrons (or holes) in the active layer in the, say, positive (negative) direction; through Coulomb scattering, momentum is transferred to the passive layer, which drives the carriers (regardless of whether these are electrons or holes) in the same direction as the momentum transfer. In doped systems where there is only one type of carrier (either electron or hole), this gives a finite drag current in the passive layer. Now, in undoped systems where a perfect electron-hole symmetry exists, there are two cases for consideration: (1) If the active layer is undoped, equal numbers of electrons and holes in the active layer will be driven in the opposite direction by the applied electric field, and the net momentum transfer is thus zero. There will be no drag regardless of what the passive layer is. (2) If the active layer is doped while the passive layer is undoped, equal numbers of electrons and holes in the passive layer will be driven in the same direction by the momentum transfer, therefore resulting in a vanishing drag current. 
The conclusion of these qualitative considerations amounts to a vanishing nonlinear 
susceptibility $\Gamma(q,\omega) = 0$, which we now proof as follows. We first make a change of the integration
variable $k' = -k$ in Eq.~(\ref{eq4}), and use time-reversal symmetry
to obtain 
\begin{eqnarray}
&&{\Gamma}(\bm{q},\omega) = \tau
\sum_{\lambda,\lambda'=\pm}\sum_{k'}
\left[-\tilde{J}_{\lambda\lambda}(\bm{k'}-\bm{q})+\tilde{J}_{\lambda'\lambda'}(\bm{k'})\right] \nonumber \\
&&\mathrm{Im}\left\{\left(1+\lambda\lambda'\mathrm{cos}\theta_-\right)\frac{n_F(\epsilon_{\bm{k'}\lambda'})-n_F(\epsilon_{\bm{k'-q}\lambda})}{\omega+\epsilon_{\bm{k'}\lambda'}-\epsilon_{\bm{k'-q}\lambda}+i0^{+}}\right\},
\label{eq5}
\end{eqnarray}
where $\theta_{-} = \phi_{\bm{k}}-\phi_{\bm{k-q}}$. This is so far general. Next we impose the symmetry requirements of
the bands about $\epsilon = 0$, i.e. $\epsilon_{\bm{k'},\lambda} = 
-\epsilon_{\bm{k'},-\lambda}$ and $\tilde{J}_{\bm{k'},\lambda} = 
-\tilde{J}_{\bm{k'},-\lambda}$, and then change the band labels as $r' =
-\lambda$, $r = -\lambda'$ in Eq.~(\ref{eq5}). Finally, using the
relation $n_F(\epsilon_{\bm{k'},-r}) = 1-n_F(\epsilon_{\bm{k'},r})$ 
valid for the undoped case $\mu = 0$,
we arrive at $\Gamma = -\Gamma$, i.e. $\Gamma(q,\omega) \equiv 0$. This result 
holds true for any type of spectrum where the two bands have a mirror 
symmetry across $\epsilon = 0$, and any bilayer system 
with one or both of the layers having such a band symmetry with zero
doping always results in an overall vanishing drag at all temperatures.   

\section{DRAG IN EXTRINSIC GRAPHENE SYSTEMS}
We now move on to the drag between finite-doped graphene layers. We
first provide the exact analytical results for the nonlinear susceptibility Eq.~(\ref{eq4})
with Fermi energy $\varepsilon_F > 0$ (in the following $x = q/k_F$ and $y =
\omega/\varepsilon_F$, and $\tilde{\Gamma} = \Gamma/(2ek_F\tau/\pi))$:
\begin{equation}
\tilde{\Gamma}(x,y) = \tilde{\Gamma}_{\mathrm{intra}}\theta(x-y)+\tilde{\Gamma}_{\mathrm{inter}}\theta(y-x),
\label{eq6}
\end{equation}
where $\theta$ is the unit step function, $\tilde{\Gamma}_{\mathrm{intra}}$ is the intraband contribution to the
nonlinear susceptibility given by the terms with $\lambda = \lambda'$ in
Eq.~(\ref{eq4}),
\begin{eqnarray}
&&\tilde{\Gamma}_{\mathrm{intra}}(x,y) = 
\frac{1}{4x}\sqrt{x^2-y^2}\left\{2\sqrt{(y+x-2)(y-x-2)}\right. \nonumber \\
&&-\sqrt{x^2-y^2}
\left[\mathrm{tan}^{-1}\left[\frac{\sqrt{(y+x-2)(y-x-2)}\sqrt{x^2-y^2}}{x^2-2-(y-2)y}\right]\right.
  \nonumber \\
&&\left.\left.+\pi\theta\left[y(y-2)-x^2+2\right]\right]\right\}\theta(2-x-y)-\left\{y\to
-y\right\}, \label{eq7}
\end{eqnarray}
and $\tilde{\Gamma}_{\mathrm{inter}}$ is the interband contribution given by
the terms with $\lambda \neq \lambda'$ in Eq.~(\ref{eq4}),
\begin{eqnarray}
&&\tilde{\Gamma}_{\mathrm{inter}}(x,y) = 
-\frac{1}{4x}\sqrt{y^2-x^2}\left\{2\sqrt{(x+y-2)(x-y+2)}\right. \nonumber \\
&&+\sqrt{y^2-x^2}
\left[\mathrm{tan}^{-1}\left[\frac{\sqrt{(x+y-2)(x-y+2)}\sqrt{y^2-x^2}}{x^2-2-(y-2)y}\right]\right.
  \nonumber \\
&&\left.\left.-\pi\theta\left[x^2-2-(y-2)y\right]\right]\right\}\theta(x+y-2)\theta(x-y+2). \label{eq8}
\end{eqnarray}
The intraband contribution correponds to electron-hole excitations in
the vicinity of the Fermi level within the conduction band, which occur
at $\omega < vq$; whereas the interband contribution corresponds to
electron-hole excitations from the valence band to the conduction band,
which occur at $\omega > vq$. Using Eqs.~(\ref{eq3}), (\ref{eq7})-(\ref{eq8}) and the 
expression for the graphene polarizability \cite{Polar}, we have
calculated numerically the drag resistivity $\rho_D$ for different values of interlayer distance $d$ and
density $n$ (Fig.~\ref{fig4}). Before we proceed to discuss our numerical results, it is 
instructive to obtain an analytical formula for the drag resistivity
under certain approximations.  To this end, we first define the standard dimensionless 
parameter for the excitation energy $u = 
y/x = \omega/vq$ in the Fermi liquid theory, which is the ratio of the phase velocity of 
the excitation $\omega/q$ to the quasiparticle velocity $v$. At low temperatures and 
with large interlayer separation, the dominant contribution to the drag conductivity Eq.~(\ref{eq3}) comes
from region with small $q$ and $\omega$, consequently the nonlinear susceptibility can
be evaluated in the limits of long wavelength $x \ll 1$ and low
energy $u \ll 1$, allowing a closed-form expression for
$\Gamma(q,\omega)$ to be extracted. The interband 
($\lambda \neq \lambda'$) contribution in Eq.~(\ref{eq4}) is in 
general smaller than the intraband ($\lambda = \lambda'$) contribution
by an order $\mathcal{O}(x^2)$, and vanishes in the limit $u \ll 1$ 
as seen from Eq.~(\ref{eq6}). This is because, in the presence of a finite Fermi level, electrons take more  
energy to transition from the valence band to the conduction band
(interband) than to transition within the conduction band
(intraband), and with a small excitation energy the channel of interband 
transition becomes inaccessible. With the above assumptions,
Eq.~(\ref{eq4}) can be evaluated as
\begin{eqnarray}
&&\tilde{\Gamma}(x,y=ux) = \tilde{\Gamma}_{\mathrm{intra}} = -4u
\left[\left(1-ux\right)\frac{t_+\theta(1-|t_+|)}{\sqrt{1-t_+^2}}\right.
 \nonumber \\
&&\left.-\left(1+ux\right)\frac{t_-\theta(1-|t_-|)}{\sqrt{1-t_-^2}}\right]
\simeq -4ux  \label{eq9}
\end{eqnarray}
where $t_{\pm} = u\pm x(1-u^2)/2$. Eq.~(\ref{eq9}) is
larger than the corresponding expression for the nonlinear susceptibility in regular 2DEG 
by a factor of $4$, due to an extra $2\times 2$ degrees 
of freedom coming from the spin and valley degeneracies in graphene, in
addition to the two sublattice degrees of freedom which give rise
to the conduction and valence bands.

The longitudinal conductivity can be obtained from the impurity-dressed current 
$\bm{J} = (\tau_{\textrm{tr}}/\tau)ev\bm{\hat{\sigma}}$ using the Kubo formula to give $\sigma_{\textrm{L}} = e^2\nu D$, where 
$\nu = 2k_F/\pi v$ is the graphene density of states and $D = 
v^2\tau_{\textrm{tr}}/2$ is the diffusion constant. This Kubo formula result is
identical with the Boltzmann theory result $\sigma_{\textrm{L}} = (e^2/\hbar) 
2\varepsilon_F\tau_{\textrm{tr}}/h$. Incidentally, for short-range impurities the transport time 
is related to the lifetime simply by $\tau_{\textrm{tr}} =  
2\tau$ due to the suppression of backscattering from impurities in graphene. 
There are two types of disorder in substrate-mounted graphene, one
being the charged impurities coming from the substrate; and the other 
being the neutral impurities intrinsic to the graphene layer itself.
In theory, the type of disorder essentially boils down to the expression of the 
transport time $\tau_{\textrm{tr}}$ in $\sigma_{\textrm{L}}$, which yield different types of functional 
dependence of the conductivity $\sigma_{\textrm{L}}$ on the carrier density. In 
experiments, the conductivity is observed to increase linearly with 
density, a fact alluding to the dominance of the charged impuritity
scattering in substrate-mounted graphene samples.  
We emphasize that this dependence on different types of disorder does not affect the expression of the 
drag resistivity as the transport time $\tau_{\textrm{tr}}$ is explicitly
canceled out between $\sigma_{\textrm{D}}$ and
$\sigma_{\textrm{L}1}\sigma_{\textrm{L}2}$. Therefore, our calculation and conclusions
apply equally to bilayer systems with substrate-mounted (where charged impurity scattering plays the more
dominant role) or suspended graphene samples (where there is only neutral impurity scattering).  

In the expression of the drag conductivity Eq.~(\ref{eq3}), the dominant contribution
of the integral comes from the region where $qd \lesssim 1$, and
for large interlayer separation $d$ satisfying $d^{-1} \ll k_F,
q_{\textrm{TF}}$ , the interlayer potential Eq.~(\ref{eq62}) can be
approximated as $U_{12} \simeq {q}/{4\pi e^2 \mathrm{sinh}(qd) \Pi_1
  \Pi_2}$. Furthermore, the denominator $\mathrm{sinh}^2(\omega/2k_B T)$ in Eq.~(\ref{eq3})
also restricts the upper limit of the $\omega$ integral to a few
$\sim k_B T$, therefore at low temperatures only small values of
$\omega/\varepsilon_F$ contribute to Eq.~(\ref{eq3}).  
As a consequence, the polarizability for doped graphene can be 
approximated by the static screening result $\Pi(q,\omega) \simeq \nu$. 
Now, using Eq.~(\ref{eq9}) for the nonlinear 
susceptibility, the drag resistivity is obtained as
\begin{equation}
\rho_{\mathrm{D}} =
\frac{h}{e^2}\frac{\pi\zeta\left(3\right)}{32}\frac{(k_B
T)^2}{\varepsilon_{F1}\varepsilon_{F2}}\frac{1}{(q_{\mathrm{TF1}}d)
(q_{\mathrm{TF2}}d)}\frac{1}{(k_{F1}d)(k_{F2}d)}, \label{eq10}
\end{equation}
where $q_{\mathrm{{TF}}} = 4e^2k_F/v$ is the Thomas-Fermi
wavenumber for extrinsic graphene \cite{gFL}. The drag resistivity 
Eq.~(\ref{eq10}), valid for low temperatures $T \ll T_F$ and high
density and/or large interlayer separation $k_F d \gg 1$, has exactly
the same form as in the regular 2DEG 
drag, exhibiting the same dependences of temperature ($\sim T^2$), interlayer separation ($\sim 1/d^4$) and
density ($\sim (n_1 n_2)^{-3/2}$). 

\begin{figure}[t]
  \includegraphics[width=5.7cm,angle=270]{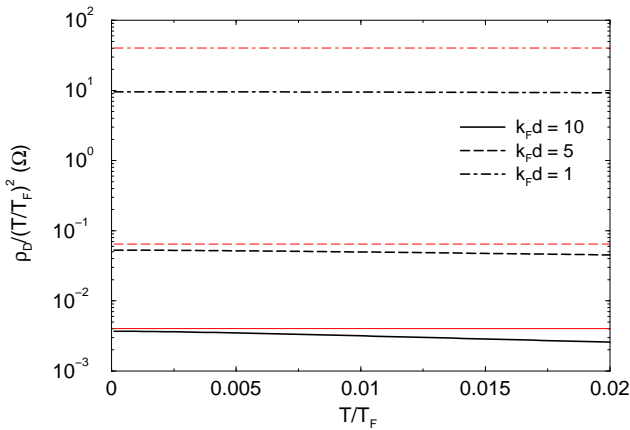}
\caption{(Color online) $\rho_{\mathrm{D}}/(T/T_F)^2$ as 
  a function of $T/T_F$ for Coulomb drag between two 
  identical extrinsic graphene sheets, with values of $k_F d = 10$ (solid lines), $5$ (dashed lines),
and $1$ (dot-dashed lines). Numerical results are indicated with bold (black) lines and analytical results 
Eq.~(\ref{eq10}) with thin (grey/red) lines. The analytical results become an increasingly accurate 
 approximation to the full numerical results with increasing $k_F d$ (i.e. increasing $n$ or $d$).}
\label{fig3}
\end{figure}

\begin{figure}[t]
  \includegraphics[width=7cm,angle=270]{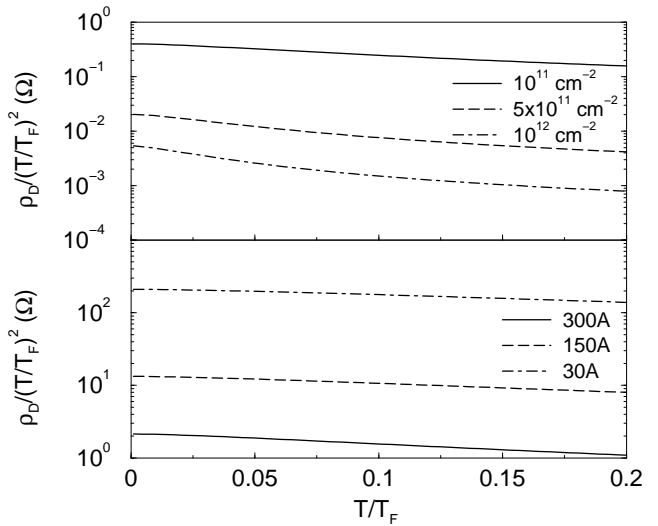}
\caption{$\rho_{\mathrm{D}}/(T/T_F)^2$ vs. 
  $T/T_F$ for higher values of $T$ up to $0.2T_F$. Upper panel: for fixed
  interlayer distance $d =
  500\AA$ and different values of density $n =
  10^{11}\mathrm{cm}^{-2}$ (solid line),
  $5\times10^{11}\mathrm{cm}^{-2}$ (dashed line),
  $10^{12}\mathrm{cm}^{-2}$ (dot-dashed line), corresponding to $T_F = 431\mathrm{K}$, 
  $963\mathrm{K}$, $1361\mathrm{K}$ respectively; lower panel: for fixed density $n =
  10^{11}\mathrm{cm}^{-2}$ and different values of interlayer distance
  $d = 300\AA$ (solid line), $150\AA$ (dashed line) and $30\AA$
  (dot-dashed line).} \label{fig4}
\end{figure}

Our numerical calculations and analytical results Eq.~(\ref{eq10}) are compared in 
Fig.~\ref{fig3}, showing that Eq.~(\ref{eq10}) becomes an increasingly accurate approximation to
the full numerical results with increasing values of $k_F d$. The fact that the exact numerical results shown in Fig.~\ref{fig3} disagree more strongly with the analytic result of Eq.~(\ref{eq10}) for smaller values of $k_Fd$ is understandable, since the analytic formula given in Eq.~(\ref{eq10}) applies only in the asymptotic $k_Fd \gg 1$ limit, and for lower carrier density and/or interlayer separation, Eq.~(\ref{eq10}) simply does not apply. In particular, for $k_Fd = 1$, the exact numerical result for Coulomb drag is a factor of $4$ larger than that given by Eq.~(\ref{eq10}). This trend of an increasing quantitative failure of the asymptotic analytic drag formula for lower values of $k_Fd$ has also been noted in the literature \cite{EHdrag} in the context of low-density hole drag in bilayer p-GaAs 2D systems. For small $k_Fd$, backscattering effects in Coulomb drag, which are unimportant for $k_Fd \gg 1$, become important. 

On the other hand, our numerical results also show that the temperature dependence of $\rho_{\textrm{D}}$ remains very close to 
$T^2$ within a wide range of temperatures for typical experimental values of $d$ and $n$ 
(e.g. $k_F d = 5$ with $n = 5\times 10^{11}\mathrm{cm}^{-2}$ and $d \simeq 400\AA$). 
The ratio of the Fermi temperature for graphene to that for regular 2DEG with parabolic spectrum (with effective mass m) 
is $T_F(\mathrm{graphene})/T_F(\mathrm{2DEG}) = mv/\hbar\sqrt{\pi n}$, so for low densities e.g. $n = 10^{11} \mathrm{cm}^{-2}$, 
$T_F = 430\mathrm{K}$ for graphene can be larger by an order of magnitude than $T_F = 42\mathrm{K}$ for GaAs 
2DEG, and the temperature dependence of $\rho_{\textrm{D}}$ for graphene drag therefore remains very closely $T^2$ up 
to about several tens of kelvin where the low temperature regime $T \ll T_F$ still remains valid, whereas for 
drag in regular 2DEG systems departure from the $T^2$ dependence of $\rho_{\textrm{D}}$ typically occurs at 
$T \lesssim 10\mathrm{K}$. The drag resistivity is calculated numerically for various values of $d$ and $n$ and higher values of temperature up to $T = 0.2T_F$ (Fig.~\ref{fig4}); $\rho_{\textrm{D}}$ is seen to grow slower and slower than $T^2$ as temperature increases. Similar dependence on temperature is also observed for drag in regular 2DEG bilayer systems before $T$ reaches $\gtrsim 0.2T_F$, beyond which $\rho_{\textrm{D}}/T^2$ starts to increase due to plasmon enhancement to the drag resistivity \cite{plasmon}. We discuss the effect of plasmon enhancement to the Coulomb drag in graphene bilayer systems in the following.


In regular 2D bilayer systems, enhancement to the drag 
resistivity due to coupled plasmon modes comes into play with
increasing temperature \cite{plasmon}. There exist two plasmon modes,
the so-called optic and acoustic modes, for which the electrons on the two 
layers move collectively in phase and out of phase, respectively, with 
each other. The energy dispersion lines for these plasmon modes lie above
the electron-hole excitation continuum (i.e., the region of $\omega$ vs. $q$ where the imaginary part of the
polarizability is non-zero, $\mathrm{Im}\Pi(q,\omega) \neq 0$) at zero temperature, and are 
not excited at low temperatures. They can be excited, however, at
higher temperatures when the electron-hole excitation continuum occupies higher
values of the excitation energy $\omega$, a consequence of the increasing gradient 
with increasing momentum $k$ in the parabolic energy dispersion 
relation. Absorption or emission of a plasmon can occur when the electron-hole excitation
continuum starts to overlap with the plasmon dispersion. 
On the other hand for graphene, because the gradient of the linear
dispersion relation is constant, increasing temperature does not 
increase the range of the possible intraband excitation energies, the
electron-hole excitation continuum being always bounded by $\omega <
vq$. This means that the plasmon excitation energy will always be out of reach from the intraband 
excitation channel at all temperatures. However, the case is different 
with the interband excitation, for which the electron-hole excitation 
continuum overlaps already at $T = 0$ with the plasmon dispersion at about \cite{Polar}
$\omega \gtrsim \varepsilon_F$. This means that plasmon-induced  
enhancement of the drag resistivity in graphene occurs, solely due to 
interband transitions, at a temperature $T \gtrsim T_F$; whereas for regular 2D systems plasmon-induced 
enhancement occurs already before $T$ reaches $T_F$ (at about $T \simeq 0.5T_F$). 



\section{CONCLUSION}
In conclusion, we have formulated the Coulomb drag problem for 
graphene bilayers. The drag resistivity is zero for intrinsic graphene. 
For extrinsic graphene, the interband contribution to the drag due to
electron-hole excitations is suppressed at low temperatures, and the
Coulomb drag is due predominantly to the intraband contribution near the
Fermi surface in the conduction band. We have obtained exact analytical results at 
$T = 0$ for both
intraband and interband contributions to the nonlinear susceptibility,
and obtained the drag resistivity numerically. We have also derived 
an approximate analytical result for the drag resistivity valid
for low temperatures, high density and/or large interlayer
separation. We find both similarities and differences for the 
graphene drag resistivity compared with that for regular 2DEG with quadratic energy spectrum. At low temperatures, 
graphene drag resistivity exhibits the same 
temperature, bilayer distance and density dependences as regular 2D
systems. For low densities $n \lesssim 10^{11}\mathrm{cm}^{-2}$, 
the low temperature regime where the $T^2$ dependence of the drag resisitivity holds 
extends by an order of magnitude that for regular 2D systems, as the Fermi temperature is higher for the 
same carrier density in graphene than in regular 2D systems. In contrast to regular 2D bilayer systems, 
there is no contribution to plasmon-induced enhancement of the drag 
resistivity due to intraband excitations, and the only 
contribution to plasmon-induced enhancement comes from interband excitations, 
which occur at temperatures $T \gtrsim \varepsilon_F$. The coupled 
plasmon modes in graphene bilayer systems can therefore be probed
experimentally with drag resistivity measurements at high enough temperatures or at low 
densities. 

Finally, we comment on the possible effect of disorder which has been argued \cite{EHB} to be important for substrate-mounted graphene monolayers in the low carrier density regime. In particular, the low-density graphene monolayers will be dominated \cite{EHB} by spatial inhomogeneities associated with electron-hole puddles induced by the charged impurities in the substrate. In the presence of such density inhomogeneity, the Coulomb drag in the low-density regime may deviate substantially from our theory based on the spatially uniform carrier density (in each layer) model. In particular, there could be large drag fluctuations including even negative drag in this low-density regime.


\section{ACKNOWLEDGEMENT}

This work is supported by US-ONR.


\begin{thebibliography}{18}

\bibitem{exp1} K.S. Novoselov, \textit{et al.}, Nature \textbf{438},
  197 (2005); Y. Zhang, \textit{et al.}, Nature \textbf{438}, 201
  (2005).
\bibitem{spectral} A. Bostwick, \textit{et al.}, Nat. Phys. \textbf{3}, 36 - 40 (2007)
\bibitem{gFL} S. Das Sarma, E.H. Hwang and W.-K. Tse, Phys. Rev. B \textbf{75}, 121406 (2007).
\bibitem{Polar} E.H. Hwang and S. Das Sarma, cond-mat/0610561; Phys. Rev. B (in press).
\bibitem{Gramila} T.J. Gramila, \textit{et al.}, Phys. Rev. Lett. \textbf{66}, 1216 (1991).
\bibitem{Smith} A.-P. Jauho and H. Smith, Phys. Rev. B \textbf{47}, 4420 (1993).
\bibitem{Oreg} A. Kamenev and Y. Oreg, Phys. Rev. B \textbf{52},
  7516 (1995).
\bibitem{Jauho} K. Flensberg, \textit{et al.}, Phys. Rev. B \textbf{52},
  14761 (1995).
\bibitem{mydrag} W.-K. Tse and S. Das Sarma, Phys. Rev. B \textbf{75}, 045333 (2007).
\bibitem{EHdrag} E.H. Hwang, \textit{et al.}, Phys. Rev. Lett. \textbf{90}, 086801 (2003).
\bibitem{plasmon} K. Flensberg and B. Y.-K. Hu, Phys. Rev. Lett. \textbf{73}, 3572 (1994); 
Phys. Rev. B \textbf{52}, 14796 (1995).
\bibitem{EHB} E.H. Hwang, S. Adam and S. Das Sarma, cond-mat/0610157; Phys. Rev. Lett. (in press). 


\end{thebibliography}
\end{document}